\documentclass[journal]{IEEEtran}
\ifCLASSINFOpdf
\else
\fi
\usepackage{amsmath}
\usepackage{amsmath,amssymb,amsfonts}%
\usepackage{amsthm}%
\usepackage{algorithm}
\usepackage{algpseudocode}
\usepackage{verbatim}
\usepackage{tabularx}
\usepackage{graphicx}
\usepackage{subfigure}
\usepackage{multicol} 
\usepackage{multirow}
\usepackage{booktabs}
\usepackage{threeparttable}
\usepackage{siunitx}
\usepackage{url}
\usepackage{bm}
\usepackage{array}
\usepackage{cite}
\usepackage[outdir=./]{epstopdf}
\usepackage[colorlinks,
    linkcolor=black,
    anchorcolor=black,
    citecolor=black
    ]{hyperref}

\usepackage{color}

\begin{document}

%
\title{Leveraging Sound Source Trajectories for Universal Sound Separation}
%
%
%
\author{Donghang Wu,~
    Xihong Wu,~
    and~Tianshu~Qu~
\thanks{Tianshu Qu is with the National Key Laboratory of General Artificial Intelligence, School of Intelligence Science and Technology, Peking University, Beijing, China, e-mail: (qutianshu@pku.edu.cn)}}

%
%

\markboth{IEEE/ACM Transactions on Audio, Speech and Language Processing,~Vol.~XX, No.~XX, XX~XXXX}%
{Shell \MakeLowercase{\textit{\emph{et al.}}}: Bare Demo of IEEEtran.cls for IEEE Journals}
%



\maketitle

\begin{abstract}
Existing methods utilizing spatial information for sound source separation require prior knowledge of the direction of arrival (DOA) of the source or utilize estimated but imprecise localization results, which impairs the separation performance, especially when the sound sources are moving. In fact, sound source localization and separation are interconnected problems, that is, sound source localization facilitates sound separation while sound separation contributes to refined source localization. This paper proposes a method utilizing the mutual facilitation mechanism between sound source localization and separation for moving sources. The proposed method comprises three stages. The first stage is initial tracking, which tracks each sound source from the audio mixture based on the source signal envelope estimation. These tracking results may lack sufficient accuracy. The second stage involves mutual facilitation: Sound separation is conducted using preliminary sound source tracking results. Subsequently, sound source tracking is performed on the separated signals, thereby refining the tracking precision. The refined trajectories further improve separation performance. This mutual facilitation process can be iterated multiple times. In the third stage, a neural beamformer estimates precise single-channel separation results based on the refined tracking trajectories and multi-channel separation outputs. Simulation experiments conducted under reverberant conditions and with moving sound sources demonstrate that the proposed method can achieve more accurate separation based on refined tracking results.
\end{abstract}


%
\IEEEpeerreviewmaketitle

\section{Introduction}
%
%
%
%
\IEEEPARstart{H}{umans} are capable of accurately localizing a sound source and separating the sound in complex and noisy environments. This phenomenon is referred to as the "cocktail party effect". The task of separating the sound of each universal source from the audio mixture is called universal sound separation.
In an actual sound separation system, the spatial information of the sound sources, especially the localization results, can further assist in sound separation. Exploring how to utilize spatial information for separation is a worthwhile research problem. Traditional sound separation algorithms that utilize spatial location information are mostly based on beamforming techniques, which suppress the signals from directions other than the target direction \cite{traditionalbaseline}. These methods often perform poorly under reverberant conditions. Additionally, they depend on the accurate estimation of the noise covariance matrix and rely on prior information about the target signal.

Recently, deep neural networks have shown
significant efficacy in sound separation tasks \cite{upit},\cite{tasnet},\cite{dprnn},\cite{convtasnet}. Utilizing spatial information in sound separation with neural networks has been investigated. The typical approach involves cascading single-channel separation with beamforming techniques like minimum variance distortionless response (MVDR) beamformers \cite{sep-beam1},\cite{sep-beam2},\cite{sep-beam3},\cite{sep-beam4},\cite{multichmap}. These methods use the separation network to generate the masks of target sound and interfering sound on each channel, and the spatial covariance matrices are then calculated to generate the beamformer filters. These methods heavily rely on the performance of single-channel separation and spatial information does not effectively aid in single-channel separation; when mask estimation of the target sound is inaccurate, the performance of the beamformer degrades significantly.

Some methods use neural networks to estimate the weights of the filters of beamformers. Luo \emph{et al.} proposed FasNet to estimate the weights of the spatial filters with multi-channel mixture signals \cite{fasnet}. Subsequently, the Transform-Average-Concatenate (TAC) mechanism is introduced in FasNet to avoid selecting a reference microphone \cite{fasnet-tac}. However, the location information of the sound source has not been utilized to assist in the separation. In \cite{adl-mvdr}, an all-deep-learning MVDR framework is proposed to estimate every component in the calculation of an MVDR filter. Liu \emph{et al.} utilizes the inter-channel phase difference (IPD) to convert the DOA into direction features to direct the separation network \cite{spatial-filter}. This kind of method requires prior knowledge of the DOA of the sound source, and its performance significantly degrades when there is a large error in DOA estimation. Some neural beamformer methods do not need to know the sources' DOA, but they require the sources to remain spatially static \cite{eabnet},\cite{taylorbeamformer}.

Another kind of method to utilize the spatial information is training the multi-channel separation networks end to end with feature extraction across the channel dimension. Lee \emph{et al.} extends the Conv-Tasnet \cite{convtasnet} and extracts the inter-channel features with 2-D TCN layers \cite{ic-conv-tasnet}. A triple-path method utilizing Squeeze-and-Excitation mechanism \cite{sq-and-ex} across the channel dimension to learn the spatial features is introduced in \cite{tri-se}. In \cite{tparn}, the dual-path transformer method \cite{dptnet} is further extended to a triple-path attentive recurrent network in which the ARN module consisting of a RNN block, an attention block and a feedforward block is applied on the inter-channel, intra-chunk and inter-chunk dimension. TF-GridNet is proposed using 2-D convolutional layers to extract spatial information, bidirectional Long Short-Term Memory networks (BLSTM) to extract intra-frame and sub-band information and multi-head self-attention to extract full-band information in \cite{tfgridnet}. SpatialNet is proposed in \cite{spatialnet} which utilizes the convolutional layers to extract spatial information followed by cross-band blocks and narrow-band blocks operated on T-F dimension and outperforms TF-GridNet with fewer parameters. The state-space model \cite{mamba} is further introduced to enhance training and inference speed in \cite{spatialnet-mamba}. However, such methods only exploit spatial information implicitly by leveraging the multi-channel information and do not effectively utilize the DOA information of the sound source. Besides, these methods are only validated under stationary sound source scenarios.

In fact, humans can effectively utilize spatial information in complex environments to simultaneously localize and separate sound sources. The localization results can aid in sound source separation, and the separation results can also assist the localization. Localization and separation are mutually supportive processes. Some studies have explored how to leverage both localization and separation simultaneously. In \cite{cone-of-silence}, the spatial area is initially divided into several regions, the sound of sources in each region is separated, and the boundaries of these regions are iteratively refined through binary search based on the iterative separation results. However, this method relies on the assumption that the sound source is stationary throughout the temporal duration of the input audio segment. Some methods implement source separation through predefined spatial partitions (e.g., distance-based separation) to separate sources within designated zones \cite{distancebased1},\cite{distancebased2},\cite{distancebased3},\cite{distancebased4}. Such methods exhibit constrained precision in spatial information utilization. Furthermore, when moving sources traverse inter-region boundaries, these algorithms exhibit separation failures due to spatial ambiguity. Taherian \emph{et al.} perform localization based on multi-channel separation and used the localization results to determine the number of sound sources in the current frame, making the subsequent continuous separation more precise \cite{leveraging}. But this method does not incorporate the localization results into the separation process. LaSNet is proposed in \cite{lasnet}, which integrates localization and separation within a single network, using loss functions for both tasks to jointly optimize the network. This approach considers localization and separation as a multi-task-learning problem, leveraging only the shared information between the two tasks in the hidden layer space, without clearly defining the mechanism by which the two tasks assist each other. This method also requires the sources to remain stationary.

Most of the sound separation methods for handling moving sound sources mostly rely on dynamically estimating the location of the sound source. Some methods estimate the location of the sources with the steered response power (SRP) methods on each frame first and track the sources based on the estimated locations. The dynamic locations are utilized as steer vectors and the sound is separated based on them \cite{move1},\cite{move2}. The localization results may deteriorate under reverberant environments, which can significantly impact the separation performance. Besides, these methods do not leverage the relevant information between localization and separation. Munakata \emph{et al.} jointly perform the sound separation and localization with an inference model with time-varying spatial covariance matrices (SCMs) and introduce the temporal smoothness of SCMs to the inference model \cite{moveicassp}. However, this method relies on assumptions about the prior distribution of the data and the sound source separation and localization do not benefit from each other. 

Inspired by coarse-to-fine iterative refinement strategies in some monaural separation approaches \cite{iter1},\cite{iter2}, in this paper, a method utilizing Mutual Facilitation mechanism between sound source Tracking and Separation (MFTS) is proposed. The proposed method consists of three stages: initial tracking, mutual facilitation and neural beamforming. In the initial tracking stage, we propose an envelope-based tracking method. We first estimate the envelope signal of each sound source, which can be viewed as simplified sound separation. These estimated envelope signals then assist the sound source tracking. The initial tracking trajectory can serve as a starting clue for mutual facilitation. During the mutual facilitation stage, the separated multi-channel signal for a source is first derived from the estimated coarse trajectory, which in turn generates a refined trajectory; these two steps iteratively refine each other through alternating updates. Finally, the neural beamforming stage utilizing the refined multi-channel separation result and trajectory achieves enhanced accuracy in the single-channel source separation. 

The rest of this paper is organized as follows. Section \ref{prob} introduces the problem formulation of the target sound extraction. Section \ref{method} presents the proposed method, including the tracking and sound extraction modules, and shows how they collaborate. Section \ref{experiments} verifies the effectiveness of the method through simulation experiments. Finally, our conclusion is presented in Section \ref{conclusion}.

\section{Universal Sound Separation}
\label{prob}
In a reverberant environment, the $M$-channel signal mixture can be formulated as follows:
\begin{equation}
\begin{split}
    y_m(n)&=\sum_{c=0}^{C}x_{cm}(n)
    =\sum_{c=0}^{C}s_c(n)*h_{cm}(n),
\end{split}
\end{equation}
where $y_m(n)$ is the mixture signal of microphone $m$ at time $n$, $C$ is the number of the sources, $x_{cm}(n)$ is the signal at the microphone $m$ from the $c$-th source, $s_c(n)$ is the signal of source $c$ and $h_{cm}(n)$ is the time-varying room impulse responses (RIR) between source $c$ and microphone $m$. The goal of universal sound separation is to separate the signal 
$\mathbf{x}_{cm}$ of  sound source $c$ at reference channel $m$. The location of the sound sources is a clue that can effectively aid in sound separation. By providing the location or DOA of source $c$ as the additional input, the sound of the target source $c$ at microphone $m$ can be extracted.

The choice of reference microphone can also affect the performance of sound source separation. To eliminate the impact of reference microphone selection on separation performance for sources at different locations, 
we use First-Order Ambisonics (FOA) signals and the zeroth-order signal is selected as the separation target. The zeroth-order ambisonic signal is equivalent to the signal recorded by an omnidirectional microphone placed at the center of a spherical microphone array, thereby neutralizing the performance variations due to different source positions and reference microphone selections. Concurrently, by transforming the microphone array signals into the spherical harmonic domain, the algorithm achieves the decoupling of the spatial and frequency information and provides an elegant framework for spatial character exploitation \cite{foa}.
\begin{figure*}[htb]
\includegraphics[width=18.0cm]{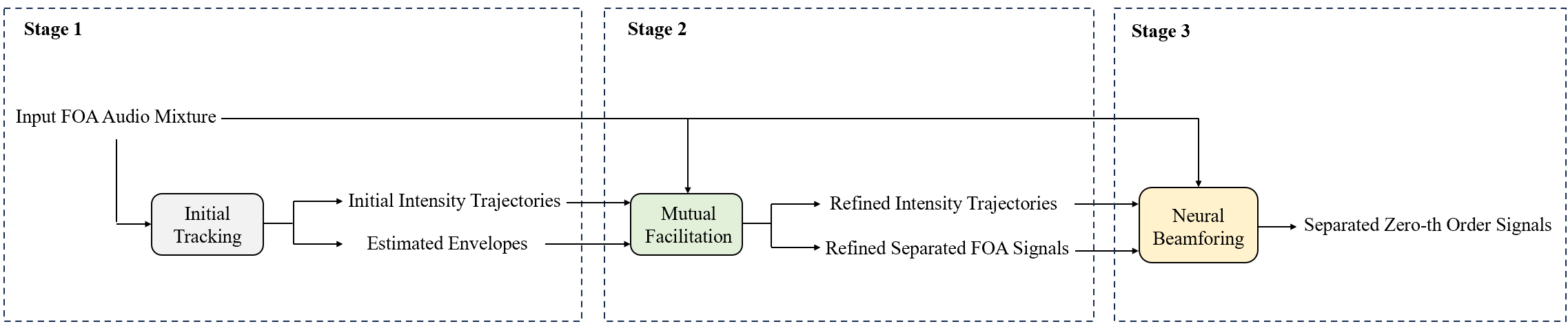}
\caption{The overall system of proposed method.}
\label{fig:overall}
\end{figure*}
\begin{figure*}[htb]
\includegraphics[width=18.0cm]{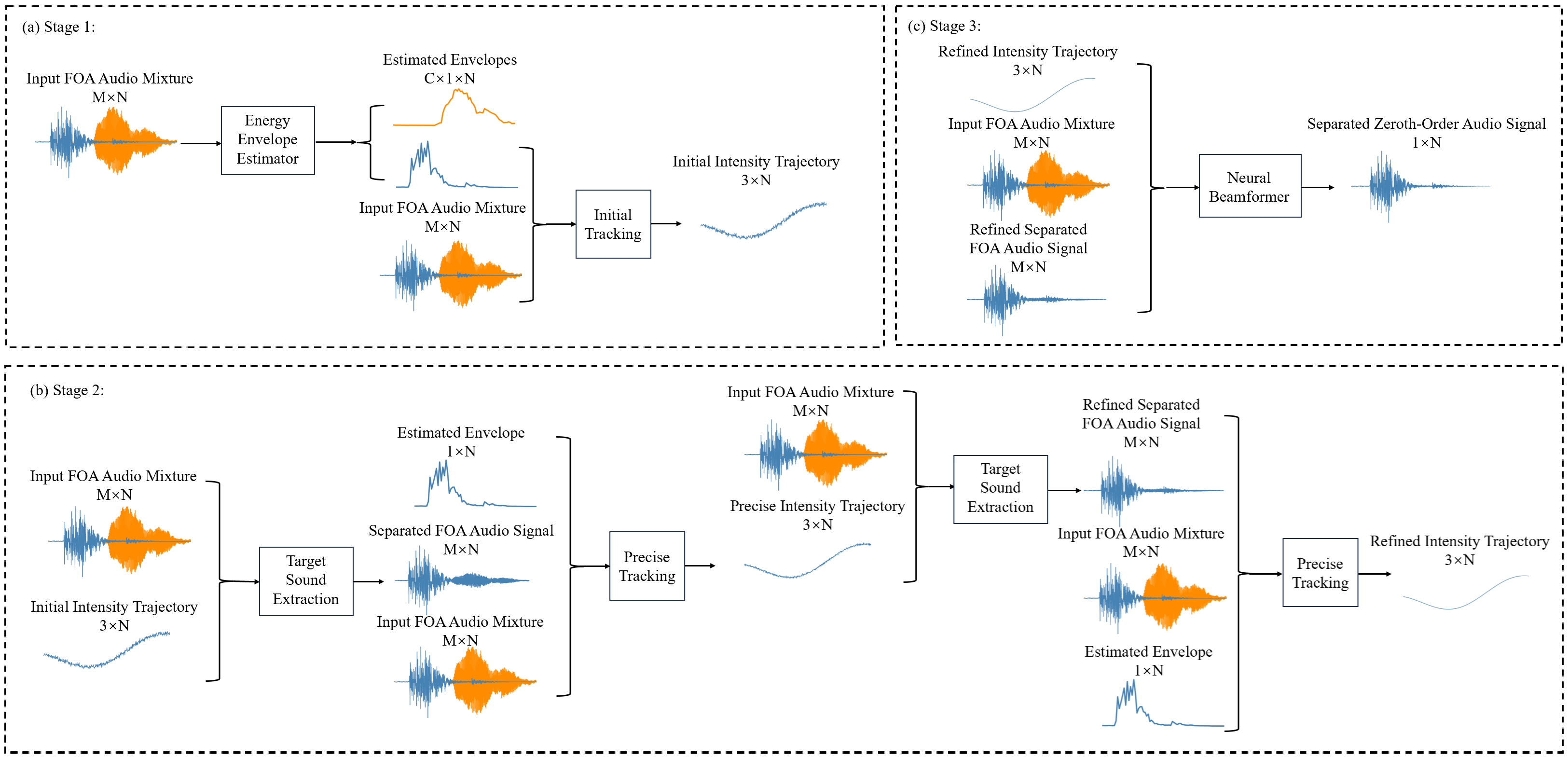}
\caption{The schematic diagram of each stage. (a) Stage 1: the initial tracking module. (b) Stage 2: the mutual facilitation module. (c) Stage 3: the neural beamforming module.}
\label{fig:structure}
\end{figure*}

\section{Method}
\label{method}
The overall schematic diagram of our proposed method is shown in Fig. \ref{fig:overall}. The proposed method can be divided into three stages. In the first stage, we perform initial tracking of all the sound sources based on the estimated envelope signal of each source. These initial trajectories might not be highly precise. In the second stage, a mutual facilitation process for source tracking and target sound extraction is performed. Specifically, the initial trajectory is used as a starting clue and input together with the mixture signals into the target source extraction network to extract the multi-channel sound corresponding to the trajectory. The separation result is subsequently input to precise tracking network and a precise tracking result based on the cleaner signal is generated. After that, this precise trajectory is then used as a clue to derive a refined multi-channel separation result based on which a refined trajectory can be then derived. In the third stage, a neural-beamformer utilizes the refined separation and tracking results to produce the single-channel sound separation signal. Fig. \ref{fig:structure} shows the schematic diagram of each stage. The training mechanisms for the above networks are described as follows. 

\begin{figure*}[t]
\begin{minipage}[b]{0.9\linewidth}
  \centering
  \centerline{\includegraphics[width=15cm]{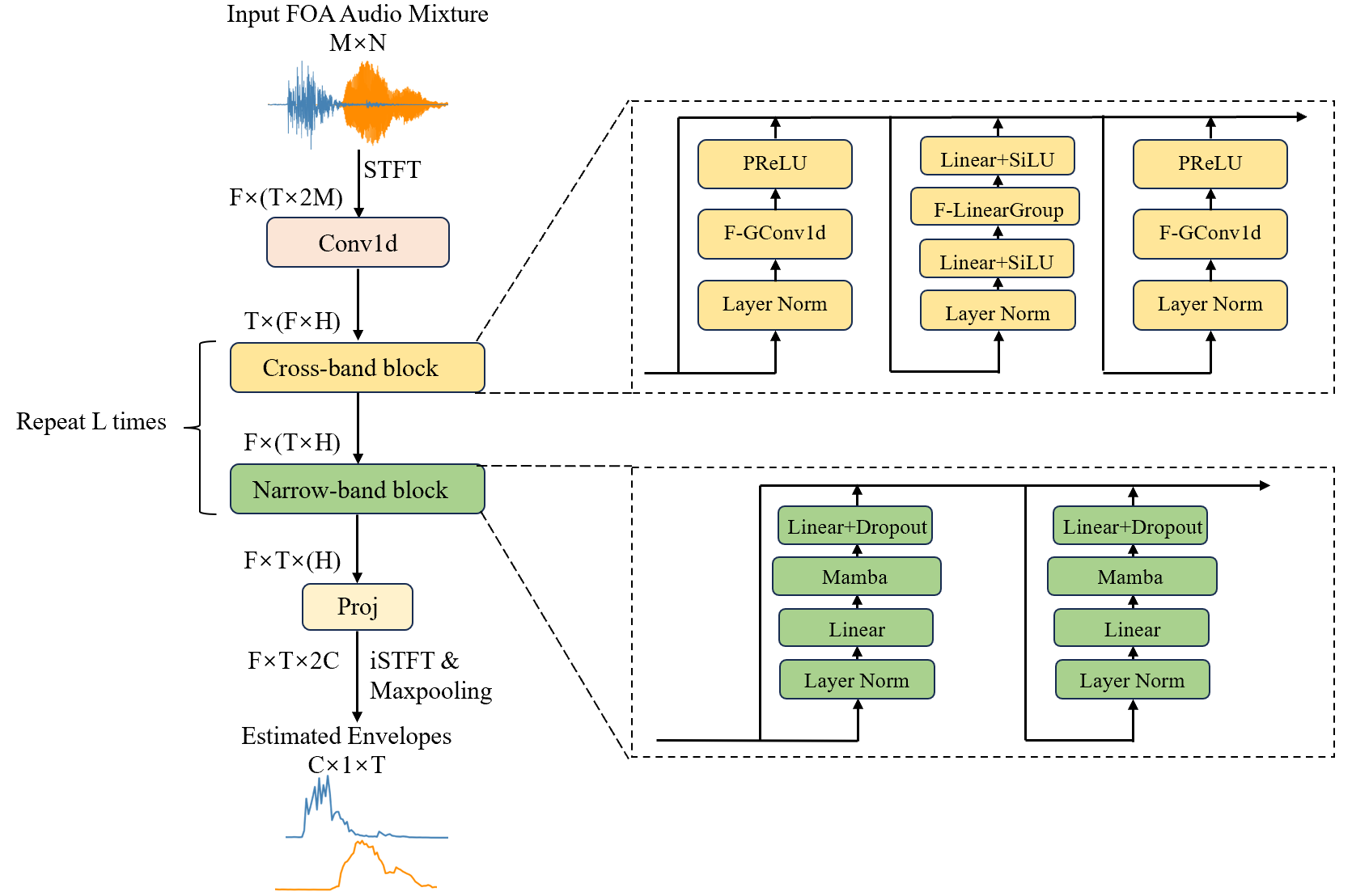}}
\end{minipage}
\caption{The structure of modified SpatialNet for envelope estimation.}
\label{fig:spatialnet-env}
\end{figure*}
\subsection{Initial tracking}
The initial tracking module takes mixture signals as input and outputs the trajectories of the sources. We first estimate the envelope signal of each sound source in the mixture. This envelope signal is then input to the tracking network along with the audio mixture signal. Concurrently, the tracking network's output is multiplied by the envelope signal to compute the loss.
\subsubsection{Envelope estimator}
The groundtruth envelope for the signal of the $c$-th source is derived through a max-pooling layer applied to the absolute value of the single-channel zeroth-order FOA signal. The max-pooling layer operates by extracting frame-wise peak values. The envelope of the $c$-th source can be denoted as $
\mathbf{\hat{E}}^\mathrm{trg}_c \in \mathbb{R}^{T\times 1}$, where $T$ is the number of frames. The full-length envelope signal of the $c$-th source, which temporally aligns with the audio mixture, is obtained by linearly interpolating $\mathbf{\hat{E}}^\mathrm{trg}_c$, which is denoted as $\mathbf{E}^\mathrm{trg}_c \in \mathbb{R}^{N\times 1}$, where $N$ is the number of temporal samples of input audio signal. $N=f_s\cdot L$, where $f_s$ is the sampling rate (Hz) and $L$ is the duration of the signal (seconds). The network used for estimating the envelopes in this study is SpatialNet \cite{spatialnet}, which is a network developed for sound separation. SpatialNet accepts multi-channel mixed signals as input and outputs single-channel separated signal for all sound sources. We select this network as the task of estimating the envelope of each sound source from the audio mixture can be considered as a simplified version of sound separation. The structure of SpatialNet for envelope estimation is shown in Fig. \ref{fig:spatialnet-env}. The mixed signal is transformed to T-F domain by Short Time Fourier Transform (STFT), which is denoted as $\mathbf{Y}\in\mathbb{R}^{F\times T\times 2M}$, where $M=4$ is the number of orders of the FOA and $T$ is the number of frames and $F$ is the number of frequency points. The T-F domain representation first undergoes a 1-D convolutional layer to extract inter-channel features. It is subsequently processed by the Cross-band block to capture frequency-band interaction features correlated with source locations. Within the Cross-band block, the input features sequentially pass through one frequency-domain group convolutional layer, one frequency-domain group of linear layers, and another frequency-domain group convolutional layer. The features are then passed through the Narrow-band block to extract intra-band temporal information. Following the method in \cite{spatialnet-mamba}, we employ Mamba to replace the original multi-head self-attention layer and the temporal convolutional layer in SpatialNet, thereby reducing computational load. Specifically, the features sequentially undergo two Mamba layers bracketed by linear layers. Finally, the features are processed through a linear projection layer, inverse STFT (iSTFT), and a max-pooling layer to obtain the estimated envelopes, $\mathbf{\hat{E}}^\mathrm{est} \in \mathbb{R}^{C\times T\times 1}$, of $C$ separated sources. The loss function for training the envelope estimation network is Normalized Mean Square Error (NMSE), which is denoted as:
\begin{equation}
\begin{split}
    \mathcal{L}_\mathrm{env}&=\text{NMSELoss}(\mathbf{\hat{E}}^\mathrm{est}, \mathbf{\hat{E}}^\mathrm{trg}) \\ 
    &= \frac{1}{C}\sum_{c=0}^{C}(10\log_{10}\frac{\sum_{t=0}^{T}(\hat{E}^\mathrm{est}_c(t) - \hat{E}^\mathrm{trg}_c(t))^2}{\sum_{t=0}^{T}(\hat{E}^\mathrm{trg}_c(t))^2}),
\end{split}
\end{equation}
where $\mathbf{\hat{E}}^\mathrm{trg} \in \mathbb{R}^{C\times T\times 1}$ denotes the groundtruch envelopes for all the sources. Besides, utterance-level permutation invariant training (uPIT) is applied during training. The estimated envelope signals are also obtained by linearly interpolating $\mathbf{\hat{E}}^\mathrm{est}$ along the temporal dimension , which is denoted as $\mathbf{E}^\mathrm{est} \in \mathbb{R}^{C\times N\times 1}$.

\subsubsection{Initial tracking}
\begin{figure}[t]
\begin{center}
\includegraphics[width=0.75\columnwidth]{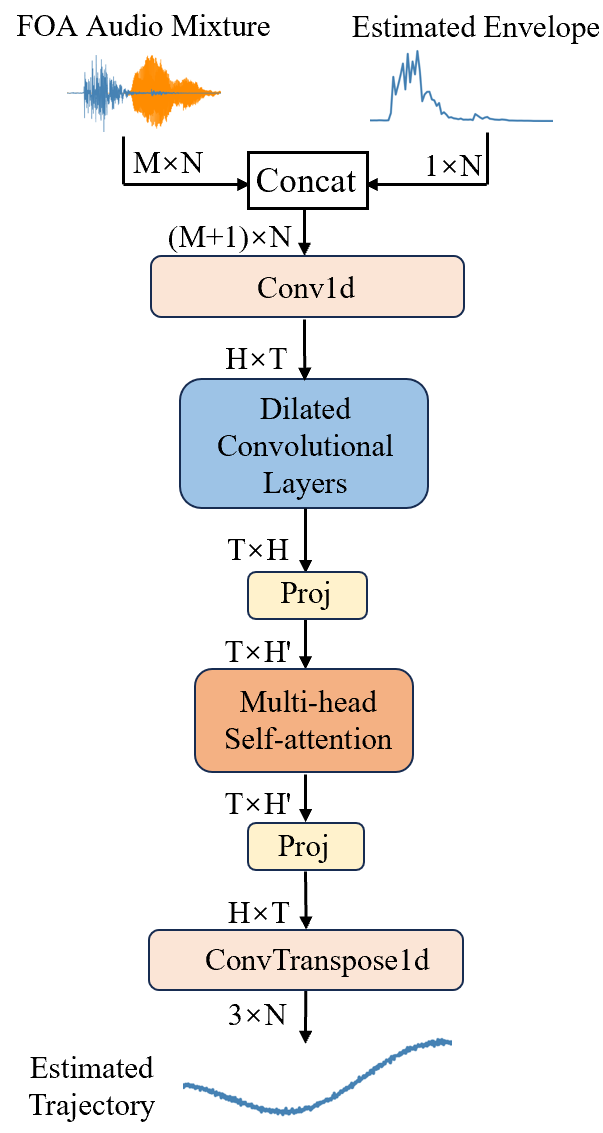}
\caption{The architecture of DCSAnet.}
\label{fg:dcsa}
\end{center}
\end{figure}

After the envelope signal of the $c$-th source $\mathbf{E}^\mathrm{est}_c$ is estimated, it is concatenated with the audio mixture and input to the tracking network to track the $c$-th source. The network used is our previously proposed DCSAnet \cite{dsca}, whose architecture is illustrated in Fig. \ref{fg:dcsa}. A 1-D convolutional layer with learnable parameters is used to encode input signals followed by dilated convolutional layers \cite{dcc}. After that a multi-head self-attention layer is utilized to capture the global temporal information. Finally, a 1-D deconvolution decoder is used to convert the features of network to temporal domain. The output of the tracking network is denoted as 
$\mathbf{traj}^\mathrm{est}_c \in \mathbb{R}^{N\times 3}$, where $N$ is the number of temporal samples of the trajectory which is equal to the length of input audio signal. The 3-D vector at each temporal sample is supposed to be the XYZ coordinates of a unit vector pointing to the sound source.

To mitigate the impact when the sound energy is low and to ensure more accurate tracking results when the sound source is active, we multiply the trajectory and the envelope signal and formulate a trajectory of intensity vectors, which we call it \textit{intensity trajectory}, denoted as $\mathbf{I\_traj}_c\in \mathbb{R}^{N\times 3}$. At each temporal sample $n$, the 3-D vector simultaneously represents the energy and direction of the sound source. The length of the vector indicates the energy of the sound source, while the direction of the vector points to the location of the sound source. This concept is similar to ACCDOA in Sound event localization and detection (SELD) tasks \cite{seld},\cite{accdoa}, but there are some differences. In ACCDOA, the vector’s length indicates the estimated likelihood of the source’s presence. In contrast, in the proposed intensity trajectory, the vector’s length reflects the source’s energy. The loss function for training the tracking network consists of two parts: the intensity trajectory loss and the differential loss:
\begin{equation}
\begin{split}
    \mathcal{L}_\mathrm{track}&=\alpha\cdot \mathcal{L}_\mathrm{traj} + \beta\cdot\mathcal{L}_\mathrm{diff}.
\end{split}
\end{equation}
We choose $\alpha=\beta=0.5$ based on experiments. The intensity trajectory loss is defined as:
\begin{equation}
\begin{split}
    \mathcal{L}&_\mathrm{traj}=\text{MSELoss}(\mathbf{I\_traj}^\mathrm{est}_c, \mathbf{I\_traj}^\mathrm{trg}_c)\\
    &=\frac{1}{N}\sum_{n=0}^{N}\|E^\mathrm{est}_c(n)\cdot \mathbf{traj}^\mathrm{est}_c(n) - E^\mathrm{est}_c(n)\cdot \mathbf{traj}^\mathrm{trg}_c(n)\|^2,
\end{split}
\end{equation}
where $\mathbf{traj}^\mathrm{trg}_c\in \mathbb{R}^{N\times 3}$ is the XYZ coordinates of the unit vector pointing to the sound source at each temporal sample. The definition of differential loss is the same as \cite{dsca}, which minimizes the error between the estimated and the target trajectory's average speed at different time scales. The differential of trajectory at time scale $d$ and sample $n$ is defined as:
\begin{equation}
\label{equa:diff}
\begin{split}
    \Delta \mathbf{traj}_{d,c}(n) = \mathbf{traj}_c(n) - \mathbf{traj}_c(n - d).
\end{split}
\end{equation}
The differential loss can be calculated as:
\begin{equation}
\label{equa:diffloss}
\begin{split}
    \mathcal{L}&_{\mathrm{diff},d}=\text{MSELoss}(\Delta \mathbf{traj}_{d,c}^\mathrm{est}, \Delta \mathbf{traj}_{d,c}^\mathrm{trg})\\
    &=\frac{1}{N-d+1}\sum_{n=0}^{N-d+1}\|\Delta \mathbf{traj}^\mathrm{est}_{d,c}(n)- \Delta \mathbf{traj}^\mathrm{trg}_{d,c}(n)\|^2,
\end{split}
\end{equation}
\begin{equation}
\label{equa:difflosssum}
\begin{split}
    \mathcal{L}_\mathrm{diff}=\frac{1}{D}\sum_{i=0}^{D}\mathcal{L}_{\mathrm{diff},2^i},
\end{split}
\end{equation}
where $\Delta \mathbf{traj}_{d,c}^\mathrm{est}$ and $\Delta \mathbf{traj}_{d,c}^\mathrm{est}$ respectively denote the differential of the estimated and target trajectories at time scale $d$. We choose $D=10$, which is the same as \cite{dsca}.

\begin{figure}[t]
\begin{minipage}[b]{0.9\linewidth}
  \centering
  \centerline{\includegraphics[width=8cm]{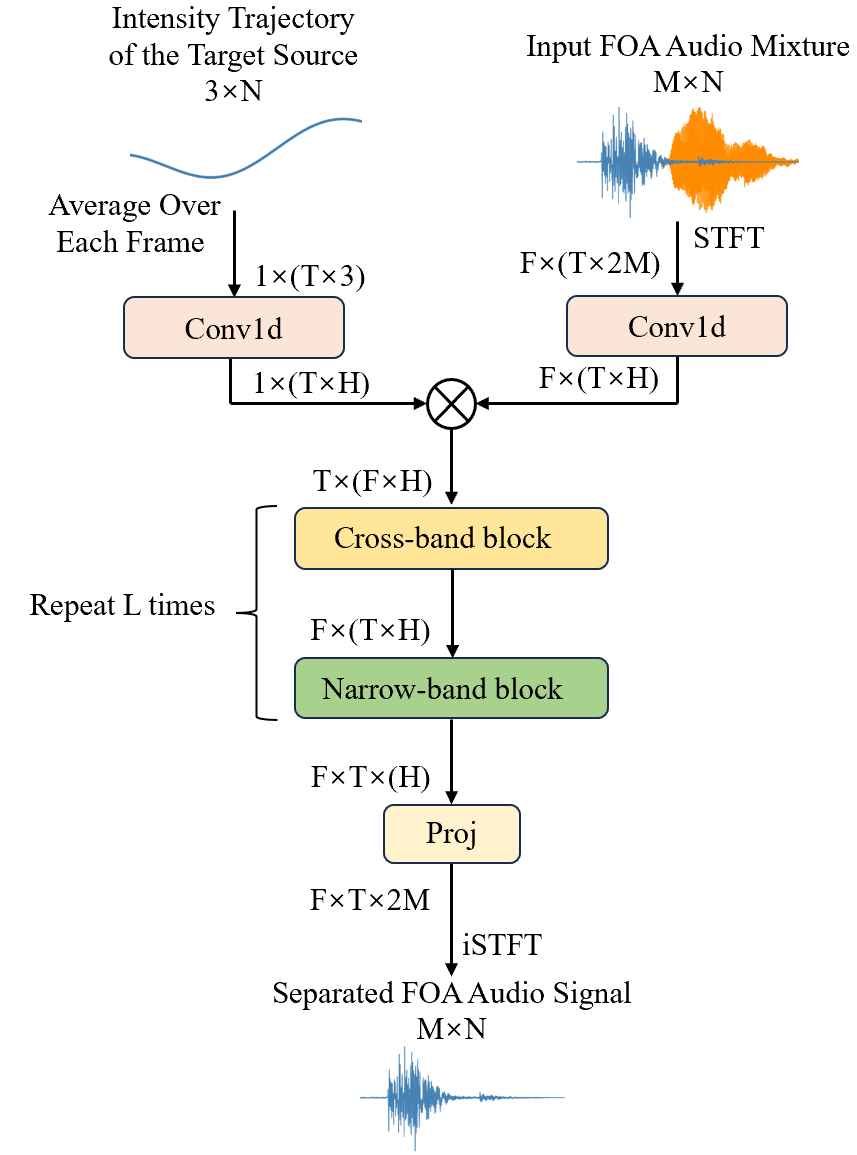}}
\end{minipage}
\caption{The structure of modified SpatialNet for target sound extraction.}
\label{fig:spatialnet-tse}
\end{figure}

\subsection{Mutual facilitation}
The mutual facilitation module includes two networks: the target sound extraction network and the precise tracking network. We will first introduce the training of these two networks and then explain the mutual facilitation process during the test phase.

\subsubsection{Target sound extraction}
The network used for target sound extraction in this study is also based on SpatialNet with Mamba \cite{spatialnet-mamba}. In this paper, we modify SpatialNet to serve as the target sound extraction network. The T-F domain representation $\mathbf{Y}\in\mathbb{R}^{F\times T\times 2M}$ of the audio mixture first undergoes a 1-D convolutional layer to obtain signal features $\mathbf{Y_f}\in\mathbb{R}^{F\times T\times H}$, where $H$ is the size of hidden state. The intensity trajectory of the target sound source is also converted into frames by averaging the intensity trajectory in one frame, which can be denoted as $\mathbf{I\_Traj}_c\in\mathbb{R}^{1\times T\times 3}$. Then the framed intensity trajectory undergoes a different 1-D convolution encoder to obtain trajectory features, which can be denoted as $\mathbf{I\_Traj\_f}_c\in\mathbb{R}^{1\times T\times H}$. The signal feature and the trajectory feature are then multiplied together and fed into subsequent cross-band and narrow-band modules. Unlike the original SpatialNet, where the network's output was the single-channel signals for all sound sources, the modified SpatialNet now outputs the FOA multi-channel signal of the target sound source. The architecture of the modified SpatialNet is shown in Fig. \ref{fig:spatialnet-tse}.

In order to establish the performance upper bound of the target sound extraction network, ensuring that its separation performance improves with the precision of the intensity trajectory, unlike the testing phase, during the training of the target sound extraction network, we do not use the output of the initial tracking module as the input. Instead, the groundtruth intensity trajectory $\mathbf{I\_traj}^\mathrm{trg}_c$ is utilized as input. Therefore, the training of the target sound extraction network is independent of the initial tracking module. The loss function for training the target sound extraction network is the negative signal-to-noise-ratio (SNR). 

\subsubsection{Precise tracking}
After training completion of the target sound extraction network, we utilize its multi-channel output signal to compute the single-channel envelope, concatenate the multi-channel signal and the calculated envelope signal along with the signal mixture and feed them into the precise tracking network. The precise tracking module is based on the cleaner separated signal thus can generate precise intensity trajectories. The network structure and loss function for precise tracking are the same as those for initial tracking, except that the number of the input channels of the 1-D convolutional layer is set to $2M+1$ ($M$ for extracted multi-channel target signal, $M$ for the signal mixture and $1$ for the estimated envelope signal of the target sound). 

Similar to the training procedure of the target sound extraction network, during training the precise tracking network, the output of the target sound extraction network when inputting the groundtruth intensity trajectory is utilized as input, which constitutes the optimal input attainable by the precise tracking network. The purpose of this approach is also to obtain the performance upper bound of precise tracking, ensuring that its tracking performance improves with the enhancement of separation performance.

\subsubsection{Mutual facilitation}
The mutual facilitation process is utilized in the testing phase. During testing, we first estimate the envelope signals of all the sound sources based on the signal mixture, select one of them, which is denoted as $\mathbf{E}^\mathrm{est}_c \in \mathbb{R}^{N\times 1}$, and perform initial tracking based on the $\mathbf{E}^\mathrm{est}_c$ and signal mixture. The initial coarse intensity trajectory denoted as $\mathbf{I\_traj}_{\mathrm{init},c}$ is the starting clue of mutual facilitation. We feed it into the target sound extraction network and the separated multi-channel signal of the target source is extracted, which is denoted as $\mathbf{x}_{\mathrm{sep1},c}\in\mathbb{R}^{N\times M}$. The $\mathbf{x}_{\mathrm{sep1},c}$ assists the precise tracking with $\mathbf{E}^\mathrm{est}_c$ as well as the audio mixture and then a precise intensity trajectory $\mathbf{I\_traj}_{\mathrm{prec1},c}$ is generated. Based on this precise intensity trajectory, we can perform the next round of mutual facilitation, thus the refined multi-channel signal $\mathbf{x}_{\mathrm{sep2},c}$ and intensity trajectory $\mathbf{I\_traj}_{\mathrm{prec2},c}$ can be generated. It is worth mentioning that, we do not change the estimated envelope signal $\mathbf{E}^\mathrm{est}_c$ during this procedure, as the first extracted multi-channel target sound $\mathbf{x}_{\mathrm{sep1},c}$ may not be accurate enough, and the envelope signal calculated based on it can do harm to the following precise tracking. Besides, we also find that the initially estimated envelope signal proves sufficiently accurate, and updating it in later stages yields no performance benefits while increasing redundancy. This mutual facilitation process is performed for each sound source based on the estimated envelopes.
\begin{figure}[t]
\begin{minipage}[b]{1.1\linewidth}
  \centering
  \centerline{\includegraphics[width=9cm]{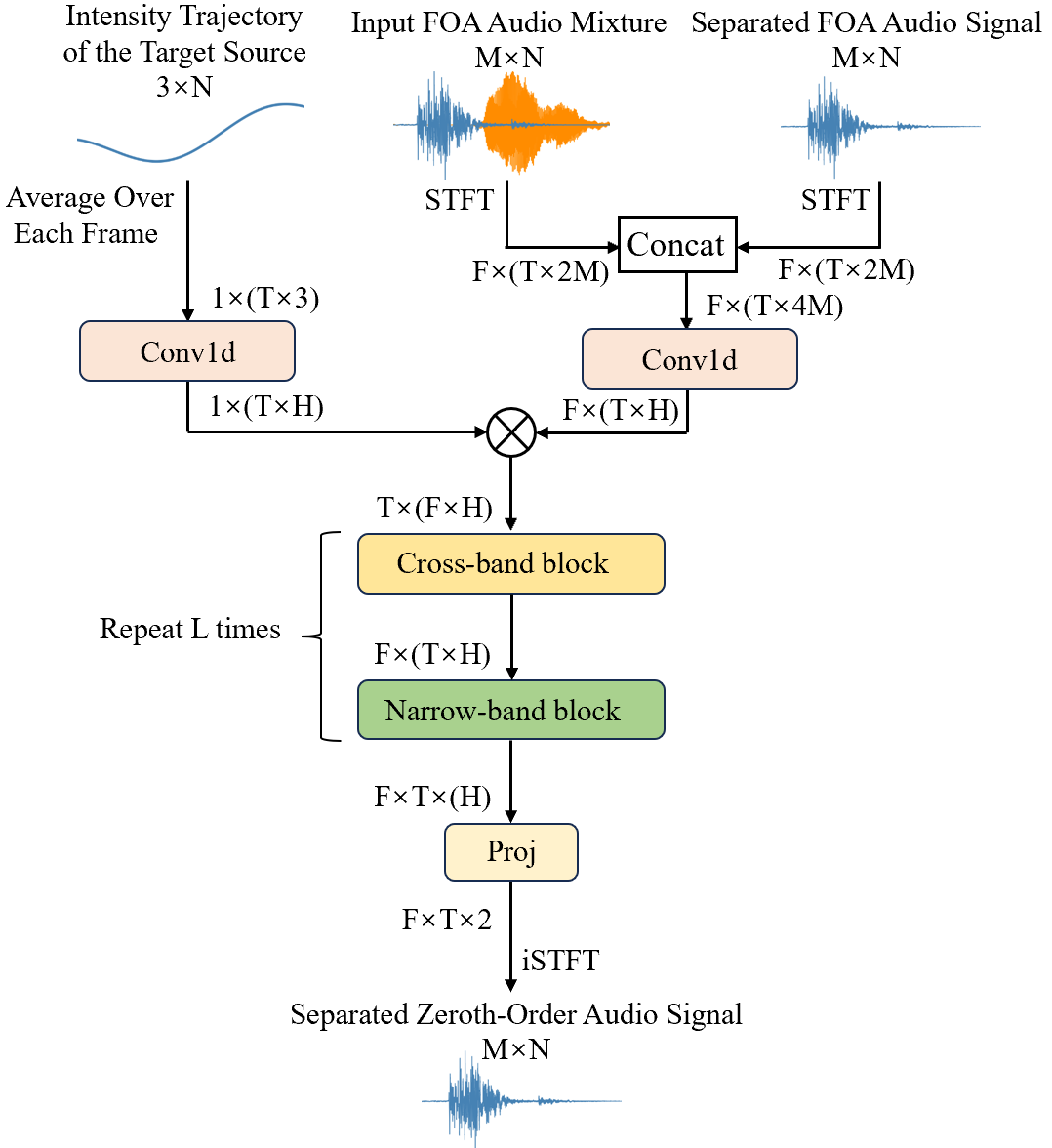}}
\end{minipage}
\caption{The structure of modified SpatialNet for the neural beamformer.}
\label{fig:spatialnet-bf}
\end{figure}

\subsection{Neural beamforming}
The neural beamforming is the final stage of our proposed method, unlike the target sound extraction and precise tracking module, it is trained after the initial tracking and mutual facilitation. 

The neural beamformer is trained on the refined multi-channel signal $\mathbf{x}_{\mathrm{sep2},c}$ and intensity trajectory $\mathbf{I\_traj}_{\mathrm{prec2},c}$, as well as the signal mixture to utilize the multi-channel information and generate a cleaner single-channel signal. In this paper, the zeroth-order signal of the FOA signal is selected as the signal of the reference channel. The network of the neural beamformer is also based on SpatialNet with Mamba. Its structure is shown in Fig. \ref{fig:spatialnet-bf}. The mixed signal and the multi-channel separated signal are transformed to T-F domain through STFT, and then they are concatenated along the channel dimension, which is denoted as $\mathbf{Y}_\mathrm{cat}\in\mathbb{R}^{F\times T\times 4M}$. The rest of the network structure of this neural beamformer is the same as that for target sound extraction, except that the number of input channels of the 1-D convolutional encoder for audio input is set to $4M$ ($2M$ for extracted multi-channel target signal, $2M$ for signal mixture), and the number of output channels of the 1-D deconvolutional layer is reduced to $2$ . The neural beamformer still multiplies the framed intensity trajectory and the time-frequency domain signal after the encoders. The loss function of training the neural beamformer follows the one of target sound extraction network.

\section{Experiments}
\label{experiments}
\subsection{Experimental setup}
\subsubsection{Dataset generation}
Our approach is validated using a simulated dataset sourced from FSD18k \cite{fsd18k}. The dataset comprises 42 hours of signals sampled at 16kHz for training, 4 hours for validation and 4 hours for testing. Room dimensions are randomly selected from a range of $(3\times 3\times 3) m^3$ to $(10\times 10\times 6)m^3$, with reverberation times $T60$ randomly assigned between $0.2$s and $1.0$s. The microphone array is configured as an Eigenmike microphone array with 32 channels \cite{eigenmike}. The captured signals are then converted into FOA format.

The method for generating trajectories of the sound sources is same as that outlined in \cite{cross3d}. Initially, two points are randomly selected within the room: a starting point $\mathbf{p_0}=(p_{x0}, p_{y0}, p_{z0})$ and an endpoint $\mathbf{p_N}=(p_{xN}, p_{yN}, p_{zN})$. Connecting these two points establishes the fundamental trajectory line. Subsequently, sinusoidal trajectories with random frequencies $\mathbf{\omega}=(\omega_x, \omega_y, \omega_z)$ and amplitudes $\mathbf{A}=(A_x, A_y, A_z)$ are added along each axis. The number of oscillation periods is constrained to not exceed $2$. The position of the sound source at temporal sample $n$ can be calculated as:
\begin{equation}
\label{equa:moving}
\begin{split}
\mathbf{p_n}=\mathbf{p_0} + \frac{n}{N}(\mathbf{p_N}-\mathbf{p_0}) + \mathbf{A}\cdot \sin(\mathbf{\omega}n),
\end{split}
\end{equation}
where $N$ is the length of the trajectory. In this experiment, only the DOA of the sound sources in utilized for target sound extraction. We set the number of sound sources to $2$. The RIRs is generated based on the Image Source Method \cite{image} with the gpuRIR Python library \cite{gpurir}. Besides, additive Gaussian white sensor noise is added to audio mixtures with a uniformly sampled in the range of $[20,30]$ dB.

\subsubsection{Network configuration}
The SpatialNet utilized consists of $8$ blocks \cite{spatialnet}. The hidden unit size in each block is $96$. The kernel size of the 1-D convolutional encoders is $5$. The kernel size of the frequency-dimension group convolution is $3$. The group numbers are $8$. In Mamba, the dmodel parameter is set to $32$, state is set to $16$, dconv is set to $4$, and expand is set to $2$. The kernel size and stride for the max-pooling layer for calculating the envelope is $256$ samples (16 ms) and $128$ samples (8 ms) respectively. The STFT window length is $256$ samples (16 ms), the hop size is $128$ samples (8 ms), and the number of frequency bins is $129$. Each network in this paper is trained with Adam optimizer \cite{adam} with learning rate of $5\cdot 10^{-4}$. 

\subsubsection{Baselines}
we compare our method with Fasnet-TAC \cite{fasnet-tac}, which is a neural beamformer, and Inter-channel Conv-Tasnet (IC Conv-Tasnet) \cite{ic-conv-tasnet}. The end-to-end networks of SpatialNet with the same architecture of the network in this paper is also utilized as baseline. We also compare the results obtained using the SpatialNet architecture to train a multi-channel mapping network (MC-SpatialNet), which outputs the multi-channel signal of each source. From the multi-channel signals generated by MC-SpatialNet, we select the zero-th channel for comparison with final output of the proposed method. The code of proposed and baseline methods is available online\footnote{https://github.com/WuDH2000/MFTS}.

\subsubsection{Metrics}
The metrics utilized for evaluating the sound separation performance is SNR, signal to distortion ratio (SDR) and scale-invariant-signal-to-noise-ratio (SI-SNR). The performance of sound source tracking depends on the energy of the sound source; when the energy is low, the localization accuracy tends to deteriorate. On the other hand, the task of sound source separation focuses more on separating the source when it is active. Therefore, good localization results should aim for greater accuracy when the sound source is active, i.e., when the energy of the source is high. Based on this, we propose an Energy-Weighted Root Mean Square Angular Error (EWRMSAE) as an enhancement of the traditional RMSAE. The EWRMSAE ($^\circ$) is defined as follows:
\begin{equation}
\label{equa:en}
\begin{split}
    &\text{EWRMSAE} = \\ &\sqrt{\frac{\sum_{n=0}^{N}(E^\mathrm{trg}(n))^2\cdot (\text{AE}(\mathbf{traj}^\mathrm{est}(n), \mathbf{traj}^\mathrm{trg}(n))) ^2}{\sum_{n=0}^{N}(E^\mathrm{trg}(n)) ^2}},
\end{split}
\end{equation}
where $AE(\cdot)$ calculates the absolute value of the angular error between the estimated and target trajectory at each temporal sample $n$. EWRMSAE assigns more weights to angular errors when sources have higher energy, while the weights of the angular errors when sources have lower energy are less.

\begin{table}[t]
  \caption{The separation performance of different methods.}
  \renewcommand\arraystretch{1.4}
  \tabcolsep=0.11cm
   \centering
  \begin{tabular}{>{\raggedright\arraybackslash}p{2cm} >{\centering\arraybackslash}p{1.5cm} >{\centering\arraybackslash}p{1.5cm} >{\centering\arraybackslash}p{1.5cm}}
    \toprule
    \textbf{Method} &{SNR} & {SI-SNR} &{SDR}
                            \\
    \midrule
    FasNet-TAC & 10.81 & 10.03 & 11.71 \\
    IC Conv-Tasnet & 11.83 & 10.92 & 12.60\\
    SpatialNet& 13.83 & 13.03 & 14.47\\
    MC-SpatialNet& 12.51 & 11.64 & 13.17\\
    \midrule
    MFTS& \textbf{15.35}&\textbf{14.48}&\textbf{15.83} \\
    \bottomrule
  \end{tabular}
  \label{tab:sep}
\end{table}

\begin{figure}[t]
\begin{minipage}[b]{0.9\linewidth}
  \centering
  \centerline{\includegraphics[width=8.5cm]{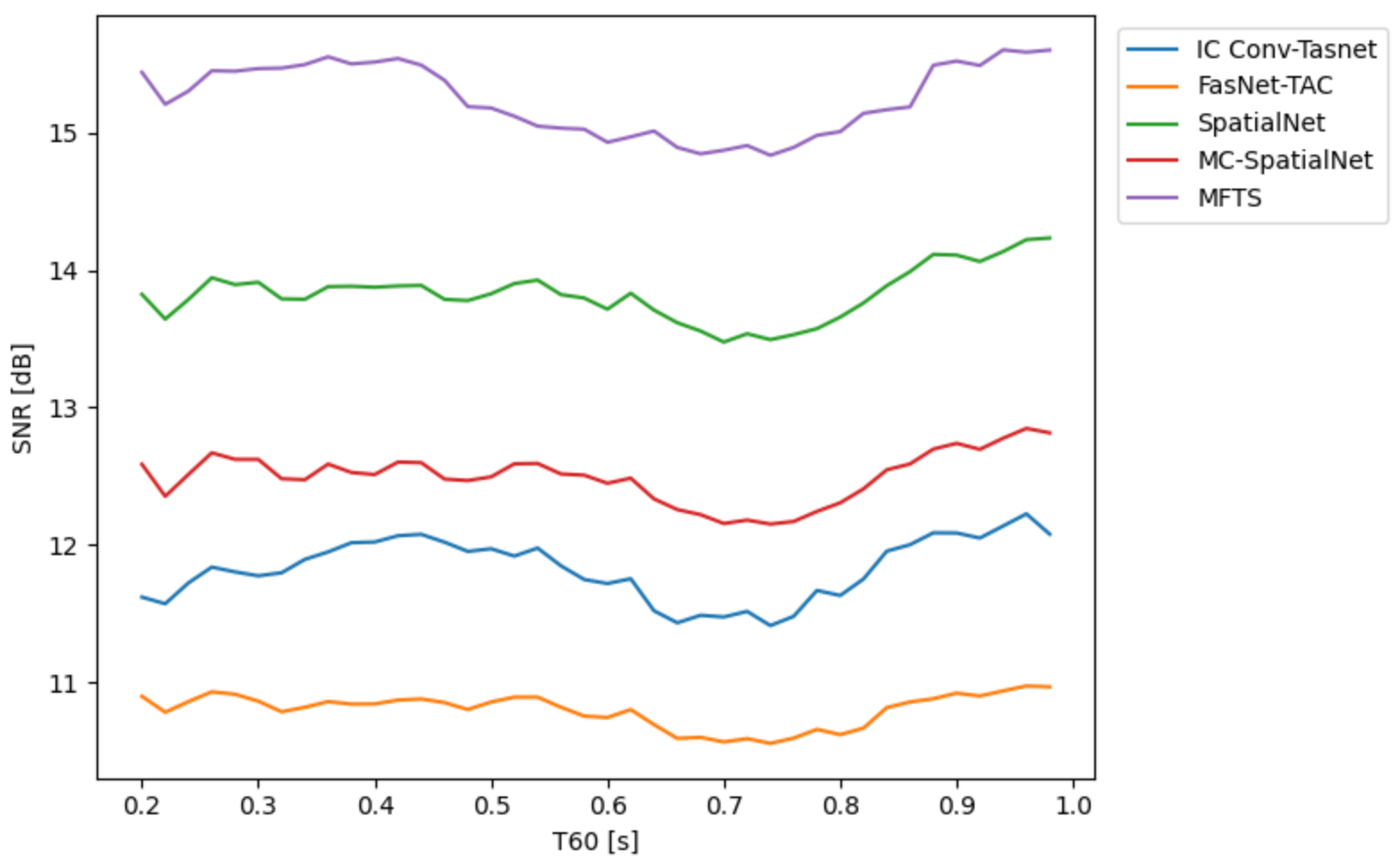}}
\end{minipage}
\caption{SNR of different methods under different T60s.}
\label{fig:sep_t60}
\end{figure}

\begin{figure}[t]
\begin{minipage}[b]{0.9\linewidth}
  \centering
  \centerline{\includegraphics[width=8.5cm]{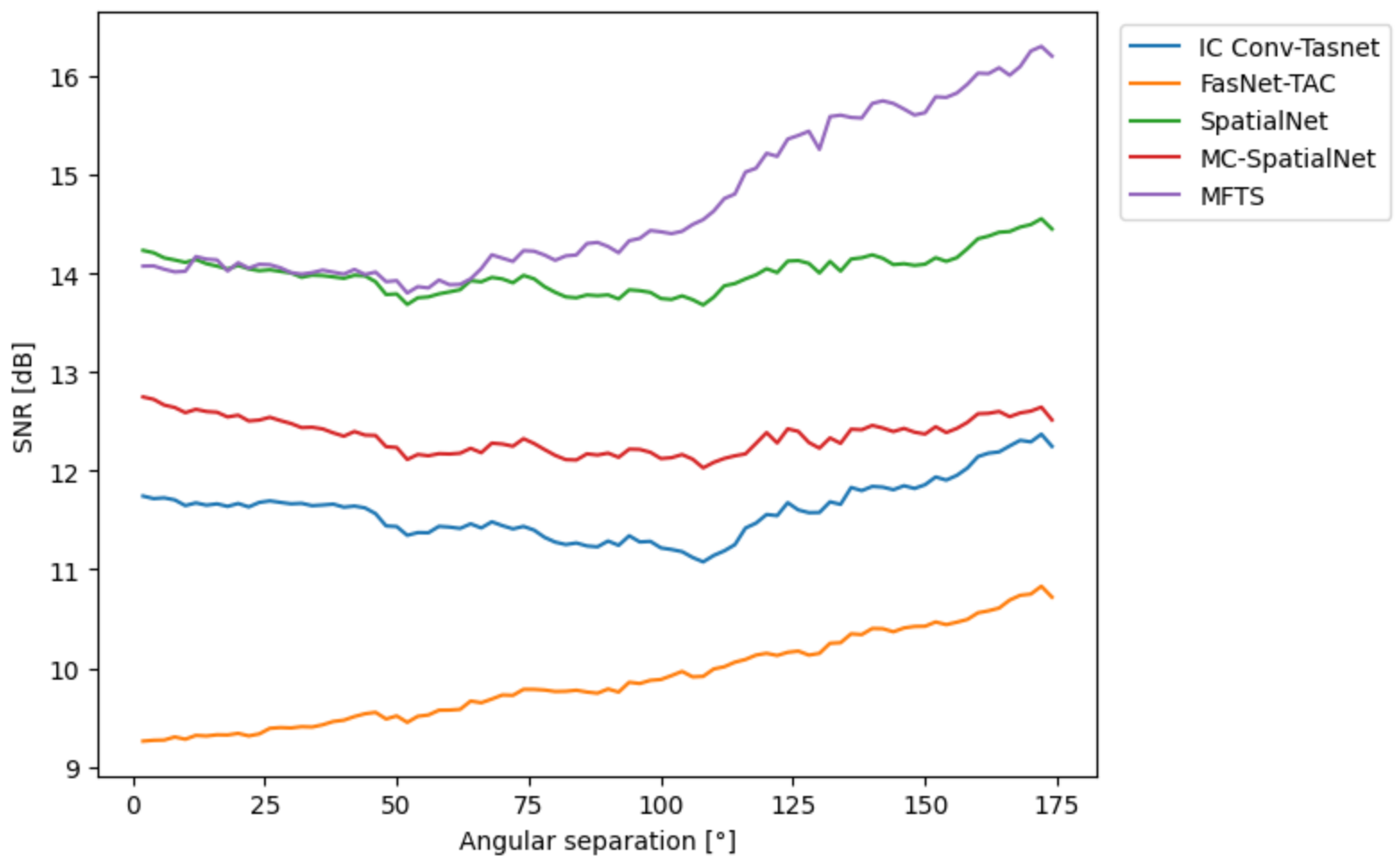}}
\end{minipage}
\caption{SNR of different methods under different angular separation between sources.}
\label{fig:sep_ad}
\end{figure}

\begin{figure}[t]
\begin{minipage}[b]{0.9\linewidth}
  \centering
  \centerline{\includegraphics[width=8.5cm]{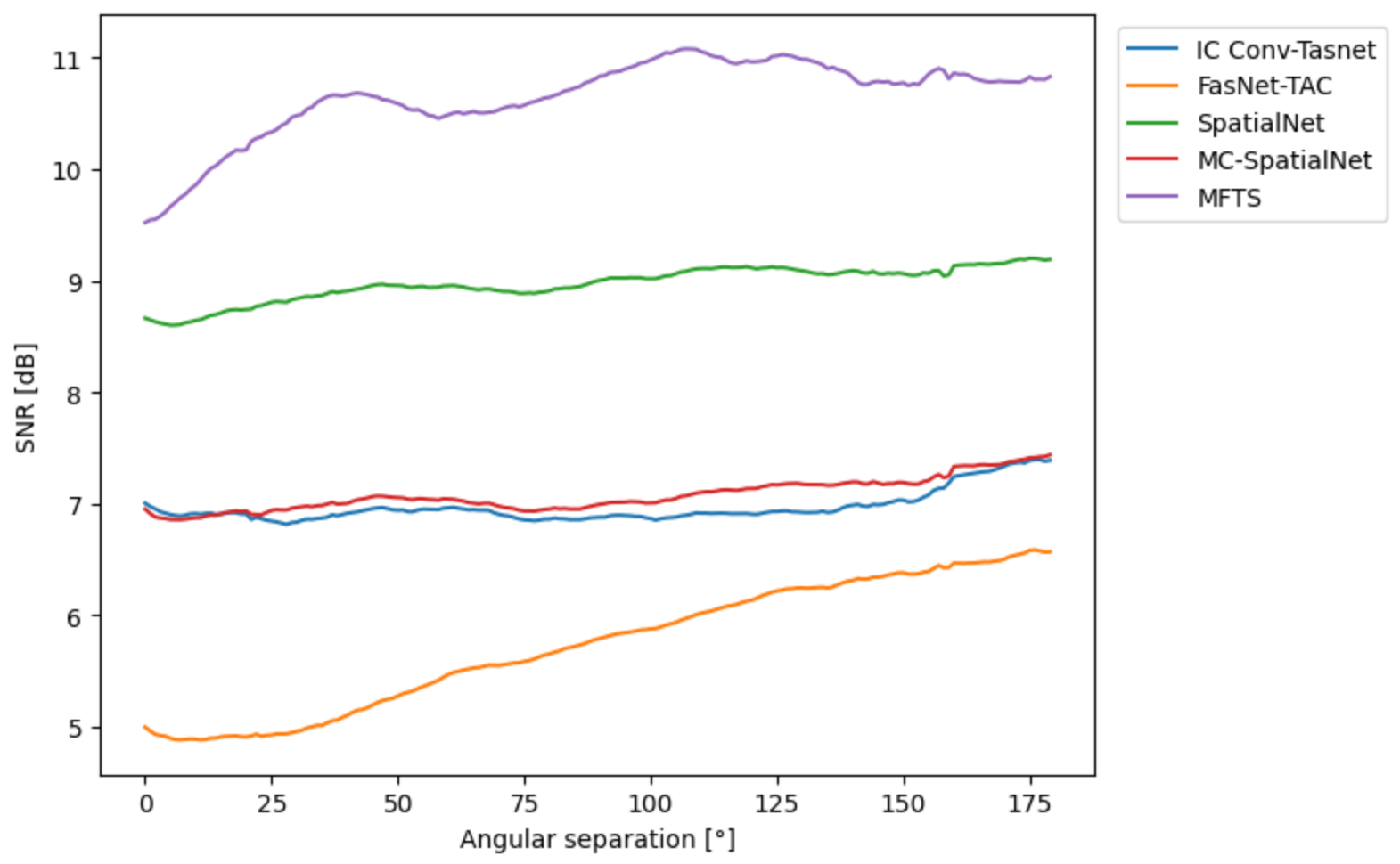}}
\end{minipage}
\caption{SNR at every 24 ms interval of different methods under different angular separation between sources.}
\label{fig:sep_ad_move}
\end{figure}

\subsection{Results}
Table \ref{tab:sep} shows each method's average performance for source separation on the test dataset. The method proposed in this paper achieves the best performance. Compared to the baseline methods, our approach explicitly tracks the sound sources and utilizes the tracking results to aid separation. The improved separation results further enhance the precision of tracking, enabling the use of refined tracking results to facilitate better separation.

To demonstrate the robustness of our proposed method, we compute the SNR values of all methods at different T60 values. Fig. \ref{fig:sep_t60} shows the results. It can be seen that our method maintains good performance under different reverberation times, and outperforms the baseline methods in all cases. 

To investigate the impact of the spatial separation between sources on sound separation performance, we concurrently evaluate the method performance under stationary source conditions and conduct a statistical analysis of separation performance across varying angular separation between two sources. The experimental configuration mirrors the moving source scenario except for fixed source positions. Fig. \ref{fig:sep_ad} illustrates the SNR value versus angular separation between sources. The $0^\circ$ angular separation between sound sources indicates that they are simultaneously active at different distances in the same direction. It can be seen that the proposed method remains functional in stationary sound source scenarios. The proposed method demonstrates superiority over all baseline approaches across various source angular separations, except when compared to SpatialNet. When the angular separation between two sources exceeds $30^\circ$, the proposed approach begins to progressively outperform SpatialNet, with the performance gap widening as the angular separation increases, particularly beyond $90^\circ$, demonstrating effective spatial information exploitation. The performance of the proposed method and SpatialNet are almost the same when the angular separation between sound sources is below $30^\circ$, which can be attributed to the localization errors of the proposed method. When the angular separation between two sound sources falls below $30^\circ$ and the sources remain stationary, the localization errors inherent in the proposed method prevent effective differentiation between the two sources. However, this limitation becomes resolvable when sound sources exhibit movement. Even when the angular separation between sound sources falls below the localization precision threshold imposed by tracking errors, our framework compensates by leveraging temporal context from intervals with sufficient spatial separation. To validate this, we quantified the SNR value and the average angular separation over every 24-ms intervals under moving-source conditions, correlating separation performance with instantaneous angular separation between sources. Fig. \ref{fig:sep_ad_move} reveals that our method surpasses all baselines in short-term separation performance under moving source scenarios regardless of instantaneous angular separation, demonstrating the critical role of motion-induced temporal dynamics in resolving spatially ambiguous cases through temporal continuity constraints.

\begin{table}[t]
  \caption{The NMSE of envelope estimation of the SpatialNet-based method and the NMF-based method.}
\renewcommand\arraystretch{1.4}
  \tabcolsep=0.11cm
   \centering
  \begin{tabular}{>{\raggedright\arraybackslash}p{2.5cm} >{\centering\arraybackslash}p{2cm}}
    \toprule
    \textbf{Method} & {NMSE}\\
    \midrule
    FastMNMF & -3.36\\
    SpatialNet & \textbf{-19.10}\\
    \bottomrule
  \end{tabular}
  \label{tab:env}
\end{table}
\begin{figure}[t]
\begin{minipage}[b]{0.9\linewidth}
  \centering
  \centerline{\includegraphics[width=8.5cm]{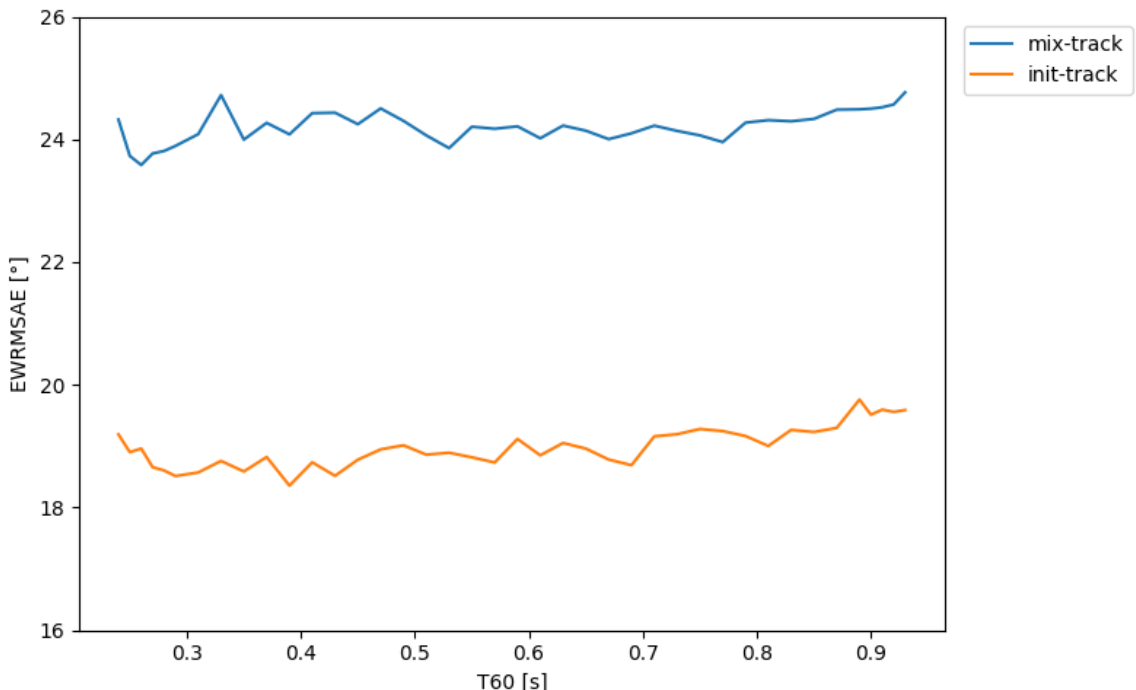}}
\end{minipage}
\caption{The EWRMSAE ($^\circ$) under different T60s.}
\label{fig:ewrmsae}
\end{figure}
\begin{figure}[t]
\begin{minipage}[b]{0.9\linewidth}
  \centering
  \centerline{\includegraphics[width=8.5cm]{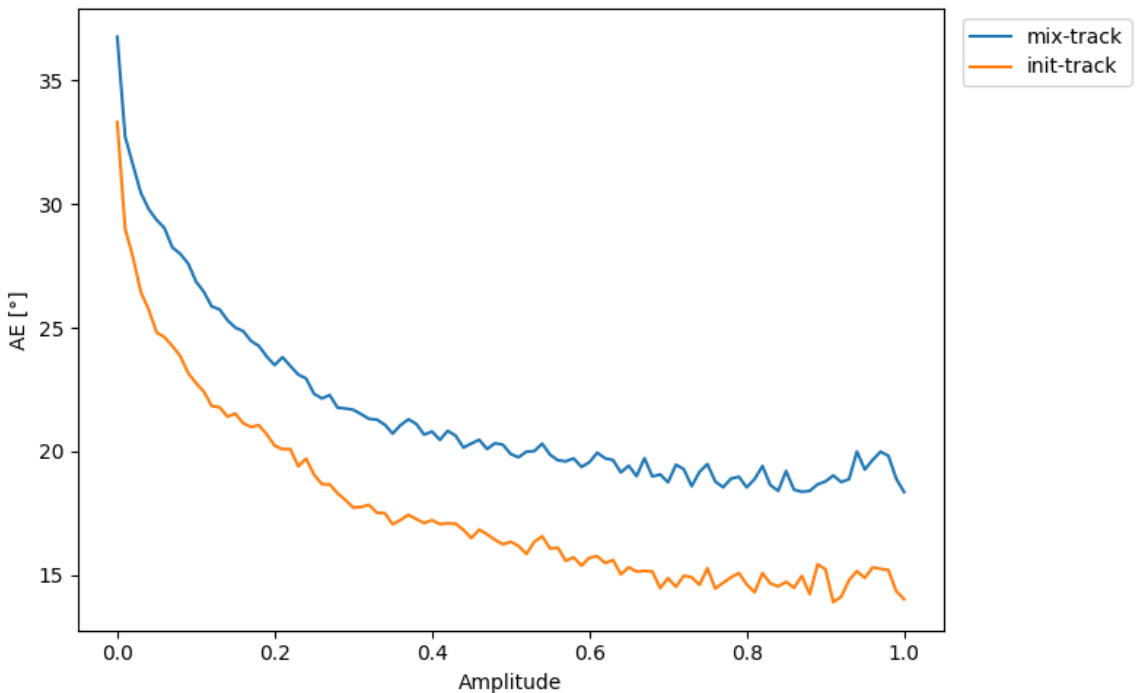}}
\end{minipage}
\caption{The angular error ($^\circ$) at individual audio samples
under different amplitudes of the envelope signals.}
\label{fig:aeea}
\end{figure}

\subsection{Ablation study}
\label{abla}
To demonstrate the effectiveness of each part of the proposed method, we perform the ablation study. The ablation study is conducted under scenarios of two moving sources.

\subsubsection{Initial tracking}
\label{ablainittrack}
We first check the effectiveness of the envelope estimation network. The proposed method based on SpatialNet is compared with the traditional signal processing method which uses FastMNMF for source separation followed by the envelope detector \cite{fastnmf}. Table \ref{tab:env} shows the result. Table \ref{tab:env} demonstrates significant superiority of the SpatialNet-based method over traditional signal processing approaches. 

To illustrate the efficiency of sound source tracking based on envelope estimation, we compare the performance of initial tracking network (init-track) to the method directly tracking the sources from the audio mixture (mix-track). Both of these two approaches utilize the same network, except that mix-track only takes the audio mixture as input, and outputs all the sources' trajectories without utilizing the envelope. The loss function for training mix-track consists of MSE between the estimated and target trajectories and the differential loss. The uPIT \cite{upit} is used during training mix-track, which means that we perform the permutation on the entire trajectory and find the order of permutation with minimum value of the loss. Fig. \ref{fig:ewrmsae} shows the EWRMSAE for both init-track and mix-track under different T60s. It can be observed that tracking based on envelope estimation results in a lower EWRMSAE across different T60 values compared to tracking without envelope information.

Additionally, we statistically analyze the envelope signal value and the angular error at every time sample, plotting the mean angular errors across different envelope signal magnitude ranges. Fig. \ref{fig:aeea} shows the results. It can be observed that when the sound source energy is low, init-track and mix-track demonstrate closely comparable error levels. This indicates that locating the sound source becomes challenging when its energy is low. Regardless of the weights assigned to moments of low energy, the network struggles to extract meaningful information from these moments. Due to init-track assigning smaller weights to the trajectory loss for low-energy moments, the differential loss carries a larger proportion in the overall loss. As a result, the network can learn more from the differential loss, which contains the motion information of the sound source. This allows the network to utilize contextual information to determine the position of the sound source at that moment. Thus the angular error of init-track is lower than mix-track. As the amplitude of the envelope signal increases, the advantage of init-track over mix-track becomes increasingly evident, which is because that init-track applies more weights to these high-energy moments than low-energy moments during one signal period thus the network can learn more from high-energy moments. Tracking the sound source based on sound source energy offers the greatest potential to improve the performance during periods of high sound source activity.
\begin{table}[t]
  \caption{The EWRMSAE ($^\circ$) of trajectories from different rounds of the mutual facilitation.}
\renewcommand\arraystretch{1.4}
  \tabcolsep=0.11cm
   \centering
    \begin{tabular}{>{\raggedright\arraybackslash}p{3cm} >{\centering\arraybackslash}p{1.5cm}}
    \toprule
    \textbf{Method} & {EWRMSAE} \\
    \midrule
    
    init-track& 19.10 \\
    MFTS-track-iter1& 14.66 \\
    MFTS-track-iter2& \textbf{14.30} \\
    MFTS-track-iter3& \textbf{14.30}\\
    \bottomrule
  \end{tabular}
  \label{tab:iter_traj}
\end{table}
\begin{table}[t]
  \caption{The separation performance from different rounds of the mutual facilitation and from the baselines.}
\renewcommand\arraystretch{1.4}
  \tabcolsep=0.11cm
   \centering
    \begin{tabular}{>{\raggedright\arraybackslash}p{2cm} >{\centering\arraybackslash}p{1.5cm} >{\centering\arraybackslash}p{1.5cm} >{\centering\arraybackslash}p{1.5cm}}
    \toprule
    \textbf{Method} & {SNR} & {SI-SNR} & {SDR} \\
    \midrule
    
    MC-SpatialNet& 12.04 & 11.15 & 12.60 \\
    Env-only & 13.41 & 12.62 & 13.94\\
    MFTS-iter1& 13.86 & 12.72 & 14.21\\
    MFTS-iter2& 14.12 & 13.09 & 14.51 \\
    MFTS-iter3& \textbf{14.14} & \textbf{13.12} & \textbf{14.53}\\
    \bottomrule
  \end{tabular}
  \label{tab:iter_sep}
\end{table}
\begin{table}[t]
  \caption{The separation performance of channel $0$ before and after neural beamforming.}
\renewcommand\arraystretch{1.4}
  \tabcolsep=0.11cm
   \centering
  \begin{tabular}{>{\raggedright\arraybackslash}p{2cm} >{\centering\arraybackslash}p{1.5cm} >{\centering\arraybackslash}p{1.5cm} >{\centering\arraybackslash}p{1.5cm}}
    \toprule
    \textbf{Method} & {SNR} & {SI-SNR} & {SDR}\\
    \midrule
    MFTS-WOBF & 14.87 & 13.92 & 15.37\\
    MFTS & \textbf{15.35} &\textbf{14.48} &\textbf{15.83}\\
    \bottomrule
  \end{tabular}
  \label{tab:bf}
\end{table}

\subsubsection{Mutual facilitation}
\label{ablamutual}
To show the efficacy of mutual facilitation mechanism, we perform the mutual facilitation three times and compare the results. Firstly, we compare the EWRMSAE values for initial tracking, the first time precise tracking, the second time precise tracking, and the third time precise tracking, represented as init-track, MFTS-track-iter1, MFTS-track-iter2, and MFTS-track-iter3, respectively. The result is shown in Table \ref{tab:iter_traj}. It can be observed that performing precise tracking on the results of multi-channel separation significantly improves sound source tracking performance. By extracting multi-channel signals based on this precise tracking and then do a further precise tracking, the tracking performance is further enhanced. Additionally, as shown in Table \ref{tab:iter_traj}, after two rounds of mutual facilitation, the tracking performance reaches its optimal value, eliminating the need for further mutual facilitation. 

Additionally, we compare the performance of target sound extraction based on the results of initial tracking, the first time precise tracking after one round of the mutual facilitation and the second time precise tracking after two rounds of the mutual facilitation. These three methods are represented as MFTS-iter1, MFTS-iter2, MFTS-iter3. The SpatialNet directly separates the multi-channel signals without trajectory inputs (MC-SpatialNet) is also compared. We also compare the method using only the magnitude of the intensity trajectory (Env-only), which is indeed the envelope signal estimation, against the approach utilizing complete intensity trajectories (both direction and magnitude), to validate the effectiveness of utilizing spatial information for the proposed method. The result is shown in Table \ref{tab:iter_sep}. It can be observed that, compared to target sound extraction based on initial tracking, the separation performance based on the precise tracking is significantly enhanced and markedly superior to the baseline method. Moreover, it can be seen that after one round of mutual facilitation, the separation results obtained in the second round of mutual facilitation reach near the optimal value. This is because the precise tracking performance approaches its optimal value, and the improvement is not significant after the second round of mutual facilitation. The Env-only method demonstrates inferior performance compared to the full trajectory method, further validating the effectiveness of spatial information utilization for separation in our framework. By analyzing the results from Table \ref{tab:iter_traj} and Table \ref{tab:iter_sep}, we choose to perform two rounds of mutual facilitation and select the separation and tracking results from the second round for the sake of effectiveness and computational simplicity.

\begin{table}[t]
  \caption{The NMSE (dB) of envelope estimation and accuracy (\%) of source count estimation under different number of sources.}
  \renewcommand\arraystretch{1.4}
   \centering
  \begin{tabular}{l c c}
    \toprule
    {\textbf{Metrics}} & {2 sources} & {3 sources} \\
    \midrule
    NMSE & -17.85 & -13.11\\
    accuracy & 92.22 & 84.06\\
    \bottomrule
  \end{tabular}
  \label{tab:unkenv}
\end{table}

\begin{table}[t]
  \caption{The EWRMSAE ($^\circ$) under different number of sources.}
  \renewcommand\arraystretch{1.4}
   \centering
  \begin{tabular}{l c c}
    \toprule
    {\textbf{Method}} & {2 sources} & {3 sources} \\
    \midrule
    init-track & 19.50 & 27.34\\
    MFTS-track-iter1 & 15.09 & 21.86\\
    MFTS-track-iter2 & \textbf{14.58} & \textbf{21.14}\\
    \bottomrule
  \end{tabular}
  \label{tab:unktrack}
\end{table}

\begin{table}[t]
  \caption{The separation performance under different number of sources.}
  \renewcommand\arraystretch{1.3}
   \centering
  \begin{tabular}{l c c c | c c c}
    \toprule
    \multirow{2}*{\textbf{Method}} & \multicolumn{3}{c|}{2 sources} &\multicolumn{3}{c}{3 sources} \\
    \cmidrule{2-7}
    ~& {SNR} & {SI-SNR} &{SDR}&{SNR} & {SI-SNR} &{SDR}
                            \\
    \midrule
    MFTS-iter1 & 12.24 & 11.20 & 12.80 & 7.56 & 6.03 & 8.00 \\ 
    MFTS-iter2 & \textbf{12.66} & \textbf{11.82} & \textbf{13.30} & \textbf{8.22} & \textbf{6.70} & \textbf{8.48} \\
    \midrule
    MFTS-WOBF & 13.35 & 12.57 & 14.13 & 8.64 & 7.25 & 9.08\\
    MFTS & \textbf{13.49} & \textbf{13.00} & \textbf{14.35} & \textbf{8.93} & \textbf{7.83} & \textbf{9.30}\\
    \bottomrule
  \end{tabular}
  \label{tab:unksep}
\end{table}

\subsubsection{Neural beamforming}
Table \ref{tab:bf} shows the separation performance on the single-channel zero-th order signal, which is equivalent to the signal recorded by an omnidirectional microphone placed at the center of a spherical microphone array, for the multi-channel separated FOA signals obtained in the second round of mutual facilitation before neural beamforming (MFTS-WOBF, we select the zero-th order to calculate metrics), as well as the separation performance after neural beamforming (MFTS). It can be observed that neural beamforming utilizes the separated multi-channel information to further enhance the separation results on the reference channel.

\subsection{Extension to an unknown number of sources}
\label{exp:unk}
As evident from the preceding descriptions, our method requires prior knowledge of the number of sound sources. This necessity arises from the envelope estimation network's requirement. The envelope estimation network needs to predefine the number of sound sources to determine its number of output channels. The task of the envelope estimation network is essentially a coarse source separation - not separating source signals themselves, but rather separating the envelopes of source signals. Numerous existing methods have achieved source separation with an unknown number of sources, such as encoder-decoder attractor (EDA) \cite{eda},\cite{tda}, recursion-based methods \cite{unkrecursive}, and approaches constrained on the maximum number of sources \cite{soundbeam}. Inspired by \cite{soundbeam}, We extend the envelope estimation network to handle an unknown number of sources under the premise of knowing the maximum number of sources $C_\mathrm{max}$. Both the network outputs and training labels have dimensions of $C_\mathrm{max}\times1\times T$. For cases where the actual source count $C < C_\mathrm{max}$, the training labels consist of $C$ ground-truth source envelopes concatenated with $C_\mathrm{max} - C$ all-zero tensor of dimension $1\times T$. The network training loss for the $c$-th channel is defined as: 
\begin{equation}
\begin{split}
    \mathcal{L}_{\mathrm{env},c}&=\begin{cases}
\text{NMSELoss}(\mathbf{\hat{E}}^\mathrm{est}_c, \mathbf{\hat{E}}^\mathrm{trg}_c)& {c<C}\\
\mathcal{L}^\mathrm{inactive}(\mathbf{\hat{E}}^\mathrm{est}_c, \mathbf{\hat{E}}^\mathrm{trg}_c)& {C \leq c \leq C_\mathrm{max}}
\end{cases}, \\ 
\end{split}
\end{equation}
where $\mathcal{L}^\mathrm{inactive}$ is defined as the loss function when $\mathbf{\hat{E}}^\mathrm{trg}_c$ is an all-zero tensor:
\begin{equation}
\begin{split}
    \mathcal{L}^\mathrm{inactive}(\mathbf{\hat{E}}^\mathrm{est}_c, \mathbf{\hat{E}}^\mathrm{trg}_c)=&10\log_{10}(\sum_{t=0}^{T}(\hat{E}^\mathrm{est}_c(t) - \hat{E}^\mathrm{trg}_c(t))^2 \\ &+ \tau\sum_{t=0}^{T}(\hat{E}^\mathrm{mix}(t))^2),
\end{split}
\end{equation}
where $\hat{E}^\mathrm{trg}_c(t)\equiv0$, $\mathbf{\hat{E}}^\mathrm{mix}$ denotes the groundtruth envelope of the zero-th order of the audio mixture and $\tau$ equals to $0.01$, which is the same as \cite{soundbeam}. The uPIT is also utilized during training. During testing, if the maximum value of an output channel in the envelope estimation network falls below $0.25$, we consider the corresponding channel to contain no active sound source. 

To validate the effectiveness of our method under an unknown number of sources, we conducted experiments under an unknown number of sources. The experimental setup mirrored the two-source scenario, except that the number of sources is randomly set to $2$ or $3$ and remained unknown to our method. Table \ref{tab:unkenv} presents the NMSE values from the envelope estimation network and the accuracy of source count estimation. The results demonstrate high estimation accuracy for both 2-source and 3-source scenarios. Table \ref{tab:unktrack} shows the source tracking performance across different number of sources and mutual facilitation rounds. Table \ref{tab:unksep} compares the multi-channel separation performance across various mutual facilitation rounds, as well as the single-channel separation results on the zero-th order signal before and after neural beamforming. While a performance degradation is observed under an unknown number of sources, our method still maintains sufficient effectiveness.

\section{Conclusion}
\label{conclusion}
In this paper, we propose a method that leverages the mutual facilitation mechanism between sound source separation and tracking to generate precise trajectories of moving sound sources for universal sound separation. Initially, we estimate the signal envelope of each source, which serves as the basis for tracking, and derive the trajectories of sources. Following this, target sound extraction is performed using the obtained trajectories, allowing for refined sound source tracking based on the separated signals. The improved tracking accuracy further enhances sound separation, creating a loop that can be repeated for increasingly accurate separation and trajectory results. Experimental results demonstrate that our proposed method outperforms baseline approaches, achieving superior performance based on the precise trajectories enabled by this mutual facilitation mechanism. This approach underscores the effectiveness of integrating sound source localization and separation to mutually enhance their respective performances.
\ifCLASSOPTIONcaptionsoff
 \newpage
\fi

\bibliographystyle{IEEEtran}
\bibliography{IEEEabrv,refs}

%






\end{document}